%% file: manuscript.tex
\documentclass[%
11pt,
onecolumn,
tightenlines,
superscriptaddress,
preprintnumbers,
nofootinbib,
amsmath,amssymb,amsthm,
physrev,
eqsecnum,tikz,
]{revtex4-2}

\input{preamble}

\usepackage{float}
\makeatletter
\let\newfloat\newfloat@ltx
\makeatother

\usepackage[english]{babel}
\usepackage[utf8]{inputenc}
\usepackage{algorithm}
\usepackage[noend]{algpseudocode}

\begin{document}

%%%%%%%%%%%%%%%%%%%%%
%%%%%%%%%%%%%%%%%%%%%
%%%%%%%%%%%%%%%%%%%%%
%%%%%%%%%%%%%%%%%%%%%

\preprint{To appear in Physical Review Materials (\url{https://doi.org/10.1103/PhysRevMaterials.6.033402}).}

\title{\Large{Comparison of simulated and measured grain volume changes during grain growth}}

\author{Xiaoyao Peng}
    \email{xiaoyaop@andrew.cmu.edu}
    \affiliation{Department of Civil and Environmental Engineering, Carnegie Mellon University}

\author{Aditi Bhattacharya}
    \affiliation{Department of Materials Science and Engineering, Carnegie Mellon University}
    
\author{S. Kiana Naghibzadeh}
    \affiliation{Department of Civil and Environmental Engineering, Carnegie Mellon University}

\author{David Kinderlehrer}
    \affiliation{Center for Nonlinear Analysis, Department of Mathematical Sciences, Carnegie Mellon University}
    \affiliation{Department of Materials Science and Engineering, Carnegie Mellon University}

\author{Robert Suter}
    \affiliation{Department of Physics, Carnegie Mellon University}
    \affiliation{Department of Materials Science and Engineering, Carnegie Mellon University}

\author{Kaushik Dayal}
    \affiliation{Department of Civil and Environmental Engineering, Carnegie Mellon University}
    \affiliation{Department of Mechanical Engineering, Carnegie Mellon University}
    \affiliation{Center for Nonlinear Analysis, Department of Mathematical Sciences, Carnegie Mellon University}

\author{Gregory S. Rohrer}
    \email{rohrer@cmu.edu}
    \affiliation{Department of Materials Science and Engineering, Carnegie Mellon University}

\date{\today}

%%%%%%%%%%%%%%%%%%%%%
%%%%%%%%%%%%%%%%%%%%%
%%%%%%%%%%%%%%%%%%%%%
%%%%%%%%%%%%%%%%%%%%%

\begin{abstract}
	
The three-dimensional microstructure of Ni, observed after five annealing intervals, was compared to simulations of grain growth using the threshold dynamics method with the assumption of capillarity as the only driving force.  A grain-by-grain comparison made it possible to identify the sources of differences between the simulation and experiment.  The most significant difference was for grains of the smallest sizes, which the simulation predicted would lose volume and disappear at a greater rate than observed in the experiment.  The loss of grains created errors in the numbers of neighbors of the remaining grains, and it was found that errors in the simulated grain volume were correlated to errors in the number of near neighbors.  While anisotropic grain boundary properties likely play a role in the differences, the size dependence of the errors suggest that it might be necessary to include a size dependence in the model for grain boundary migration kinetics. 

\end{abstract}

\maketitle

%%%%%%%%%%%%%%%%%%%%%
%%%%%%%%%%%%%%%%%%%%%
%%%%%%%%%%%%%%%%%%%%%
%%%%%%%%%%%%%%%%%%%%%

\section{Introduction}
The microstructures of polycrystalline metals and ceramics processed at high temperatures are
influenced by grain growth and this influences structure sensitive materials properties.
To better understand grain growth, analytic theories
\cite{hillert1965theory,fischer2008distribution} and, more recently, 
Monte Carlo simulations \cite{srolovitz1984Potts,srolovitz1986Potts,rollett1992Potts,peczak1993Potts}, 
cellular automata models \cite{hesselbarth1991Cellular,raabe2005Cellular}, 
molecular dynamics \cite{jhan1990molecular,upmanyu1999molecular,holm2010grain,race2014role,freitas2018free,zhang2006characterization,chen2018atomistic,chen2020atomistic}, 
vertex simulations \cite{bragg1947vertex,humphreys2000vertex}, 
phase field models \cite{chen2002phase,krill2002phase,kamachali2012phase,admal2019three,kobayashi1998vector,ito2019bayesian,moelans2008quantitative,kazaryan2001grain,bjerre2013rotation,adland2013unified,korbuly2017grain,tang2006diffuse,vanherpe2007bounding},
and other approaches \cite{farjas2007numerical,barmak2011critical,moldovan2002scaling,chen2017effect},
have been developed.

Recent studies of three-dimensional (3D) grain growth using simulation have attempted to quantify and understand topological evolution \cite{kamachali2012phase,dehoff2020topological,lazar2012complete,lazar2020distribution,eren2021constant,mason2015geometric}, volumetric growth rate \cite{yadav2018analysis}, and grain boundary energy evolution in isotropic as well as anisotropic grain growth \cite{gruber2009misorientation,sakout2020energetic}. 
There have been relatively few attempts in the past to compare the results of grain growth simulations with experiments on a grain-by-grain basis. For example, in 2003 Demirel et al. \cite{demirel2003bridging} validated an anisotropic grain growth simulation in 2D using GB curvature as a driving force and it explained 50\% of the experimental grain growth. They also showed that an isotropic simulation with 17\% explaining power had very poor matching with experimental evolution. McKenna et al. \cite{mckenna2014grain} compared the evolution of individual grains in a 3D experiment with grains in a 3D isotropic phase field simulation of grain growth in polycrystalline $\beta$-Ti. Statistical analysis revealed good agreement between experiment and simulation for grain growth kinetics. However, direct comparison of individual grains revealed a poor match in grain shapes and grain boundary widths because the simulation was unable to capture local anisotropy in grain boundary energy and mobility. 
Another study by Zhang et al. \cite{zhang2020grain} compared a phase field simulation of grain growth with 3D experimental data and assigned reduced mobilities to each grain boundary so that the simulation reproduced the experiments. 
However, the assigned grain boundary properties were independent of grain boundary crystallography in the sense that crystallographically identical boundaries had to be assigned reduced mobilities that varied by a factor of nine.

The prior work clearly shows that, unless inconsistent grain boundary properties are assigned, as in \cite{zhang2020grain}, simulations of grain growth are poor predictors of real grain growth.
However, the mechanism of how the simulations
fail has not yet been identified and the current work is aimed at
understanding this issue. We compare the outcomes of 3D grain growth
simulations with the microstructure evolution in annealed Ni \cite{hefferan2009statistics,bhattacharya2019}. For microstructures
containing 600 to 900 grains, we compare the volume changes, changes
in the numbers of near neighbors, and curvatures of grains in the
simulation and experiment. We concentrate on three places where the
simulated microstructures depart from the observations: the volume
changes for the smallest grains, grains for which the topology (number of
neighbors) is simulated incorrectly, and the grain face curvature.
While the later two observations might be connected to anisotropic
boundary properties, the first is more likely to stem from other
factors.

To simulate grain growth driven by mean curvature, we have selected
the threshold dynamics (TD) method originally introduced by Merriman,
Bence and Osher in \cite{merriman1992diffusion,merriman1994motion}. The choice of using TD to tackle
the 3D grain growth problem comes naturally
from its beneficial characteristics. First, we can use experiment
image/data directly without additional meshing. Second, the
representation of the interface is implicit, as in the phase field or
level set methods, so that topological changes are tracked
automatically. Third, TD has unconditional stability and high
computational efficiency. Finally, the isotropic formulation of the model
can be easily extended to an anisotropic formulation by replacing the
isotropic kernel with anisotropic ones assuming different anisotropic
surface energy/mobility forms, a feature we are leveraging in our
ongoing work.
	%%%%%%%%%%%%%%%%%%%%%
	%%%%%%%%%%%%%%%%%%%%%
	%%%%%%%%%%%%%%%%%%%%%
	%%%%%%%%%%%%%%%%%%%%%
	
\section{Methods}
\subsection{Threshold Dynamics}
\label{sec:TD_methods}
The threshold dynamics model assumes that capillarity is the only driving force for grain boundary motion. In TD, phases can be represented through the characteristic function (order parameter) $\mathbf{1}_{\Sigma_i}$ of each phase $\Sigma_i$. 
\begin{equation}
    \label{CharFn0}
    \mathbf{1}_{\Sigma_i}=\begin{cases}
    1, & \text{if $\bfx \in \Sigma_i$}.\\
    0, & \text{otherwise}.
  \end{cases}
\end{equation}
The evolution of the interface network is reflected by the change of the phase boundary of $\mathbf{1}_{\Sigma_i}$.
The original TD scheme from Merriman, Bence and Osher in the isotropic, two-phase setting is given as in Algorithm \ref{alg:MBO} from \cite{merriman1992diffusion}. 

\begin{algorithm}
\caption{MBO}\label{alg:MBO}
\begin{algorithmic}[1]

\State \textbf{Initialization} Given $\Sigma^0$ and time step size $\delta t$
\State \textbf{For iteration $k+1$} 
	\State \quad $\psi^k = K_{\delta t}*\mathbf{1}_{\Sigma^k}$ \Comment{Convolution/Diffusion step} 
	
	\State \quad $\Sigma^{k+1} = \{\bfx : \psi^k(\bfx)\geq \frac{1}{2} \} $ \Comment{Threshold/Sharpening step}
\end{algorithmic}
\end{algorithm}

The convolution kernel $K:\mathbb{R}^d\rightarrow\mathbb{R}$, can be any spherically symmetric kernel, and is usually chosen to be the Gaussian:
\begin{align}
    \label{unitGaussian}
    G_{\delta t}(\bfx)=\frac{1}{(4\pi(\delta t))^{d/2}}\exp(-\frac{|\bfx|^2}{4(\delta t)})
\end{align}

By this choice of kernel, the MBO scheme evolves the boundary $\partial\Sigma$ by mean curvature motion \cite{merriman1992diffusion} and the phase boundary velocity is
\begin{equation}
\label{Eq:GBV}
    \bfv(\bfx)=\mu \sigma \kappa(\bfx) \hat\bfn(\bfx)
\end{equation}
where $\kappa$ is the mean curvature, $\mu$ is the boundary mobility, $\hat\bfn$ is the boundary unit normal, and $\sigma$ is the boundary energy.

The time step size $\delta t$ is a model variable that has no effect on stability, but should be considered to reach desired accuracy. In \cite{esedoglu2015threshold}, Esedoglu and Otto introduced an extension of the two-phase MBO scheme to a general multi-phase setting, as showing in the Algorithm \ref{alg:EO15}.

\begin{algorithm}
\caption{EO}
\label{alg:EO15}
\begin{algorithmic}[1]
\State \textbf{Initialization} Given $\Sigma^0_1,..., \Sigma^0_N$ and time step size $\delta t$
\State \textbf{For iteration $k+1$} 
	\State \quad $\psi^k_i = \sum\limits_{\substack{j=1 \\ j\neq i}}^N 
	K^{i,j}_{\delta t}*\mathbf{1}_{\Sigma^k_j}$ \Comment{Convolution/Diffusion step}
	\State \quad $\Sigma^{k+1}_i = \{\bfx : \psi^k_i(\bfx)\leq \min\limits_{j\neq i}\psi^k_j(\bfx) \}$
	\Comment{Threshold/Sharpening step}
\end{algorithmic}
\end{algorithm}

The main difference between the MBO and the EO schemes is that in the EO scheme the new kernel $K^{i,j}_{\delta t}$ is designated for the boundary between $\Sigma_i$ and $\Sigma_j$. The new kernel in the EO scheme is constructed to account for non-constant $\sigma$ and $\mu$ between two phases. As mentioned in the introduction, $\sigma$ and $\mu$ play far more complex roles in grain growth than being constant. They are affected by both the grain boundary normal and misorientation between neighboring grains. Incorporating the anisotropic surface energy and mobility into $K^{i,j}_{\delta t}$ yield to an anisotropic grain growth formulation.

In this work, we focus on multi-phase isotropic simulations. We use Algorithm \ref{alg:EO15} and take the same form as in Equation (\ref{unitGaussian}) for the kernel $K^{i,j}_{\delta t}$. 
In the grain growth simulations performed in this work, Algorithm \ref{alg:EO15} makes direct use of the observed microstructures as the input by assigning one characteristic function $\mathbf{1}_{\Sigma_i}$ to each grain $\Sigma_i$, and evolves the grain boundary network by first a convolution/diffusion step and then a threshold/sharpening step.
The developed code for this study is available online at
\\ \url{https://github.com/JadeXiaoyaoPeng/GrainGrowth_TD_iso}.

%%%%%%%%%%%%%%%%%%%%%
%%%%%%%%%%%%%%%%%%%%%
%%%%%%%%%%%%%%%%%%%%%
\subsection{Model Validation}
\label{sec:TD_vali}
\begin{figure}[h!]
\centering
\includegraphics[width=0.8\textwidth]{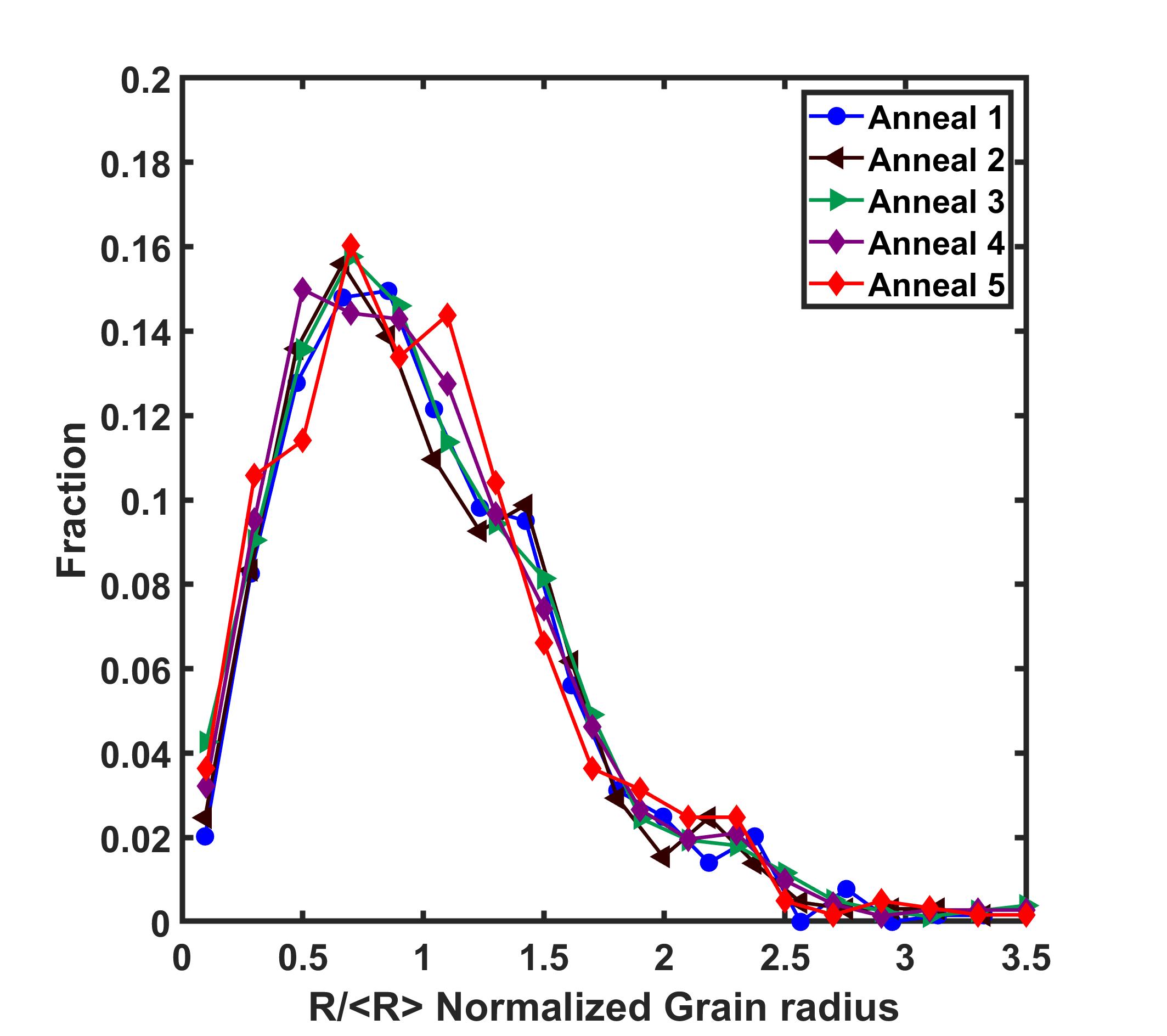}
\caption{\label{fig:gsd} Grain size distributions of simulations at different annealing states.}
\end{figure}

To validate the isotropic TD model, we compare the results of the model to the expectations from isotropic grain growth theory. For example, for normal isotropic grain growth, the resulting distribution of grain sizes is expected to be self similar \cite{burke1952recrystallization,atkinson1988overview}. 
The grain size distributions of the simulated microstructures at different anneal states are plotted in Figure \ref{fig:gsd}. 
The number of grains in each simulated anneal state from anneal state 1 to anneal state 5 are 643, 666, 775, 715 and 606, respectively.
The five curves from different time steps overlap, with some small fluctuations that result from the limited number of grains. 
Therefore, we conclude that the simulation produces self similar grain size distributions once the times scale is considered.

\begin{figure}[h!]
\centering
\includegraphics[width=0.6\textwidth]{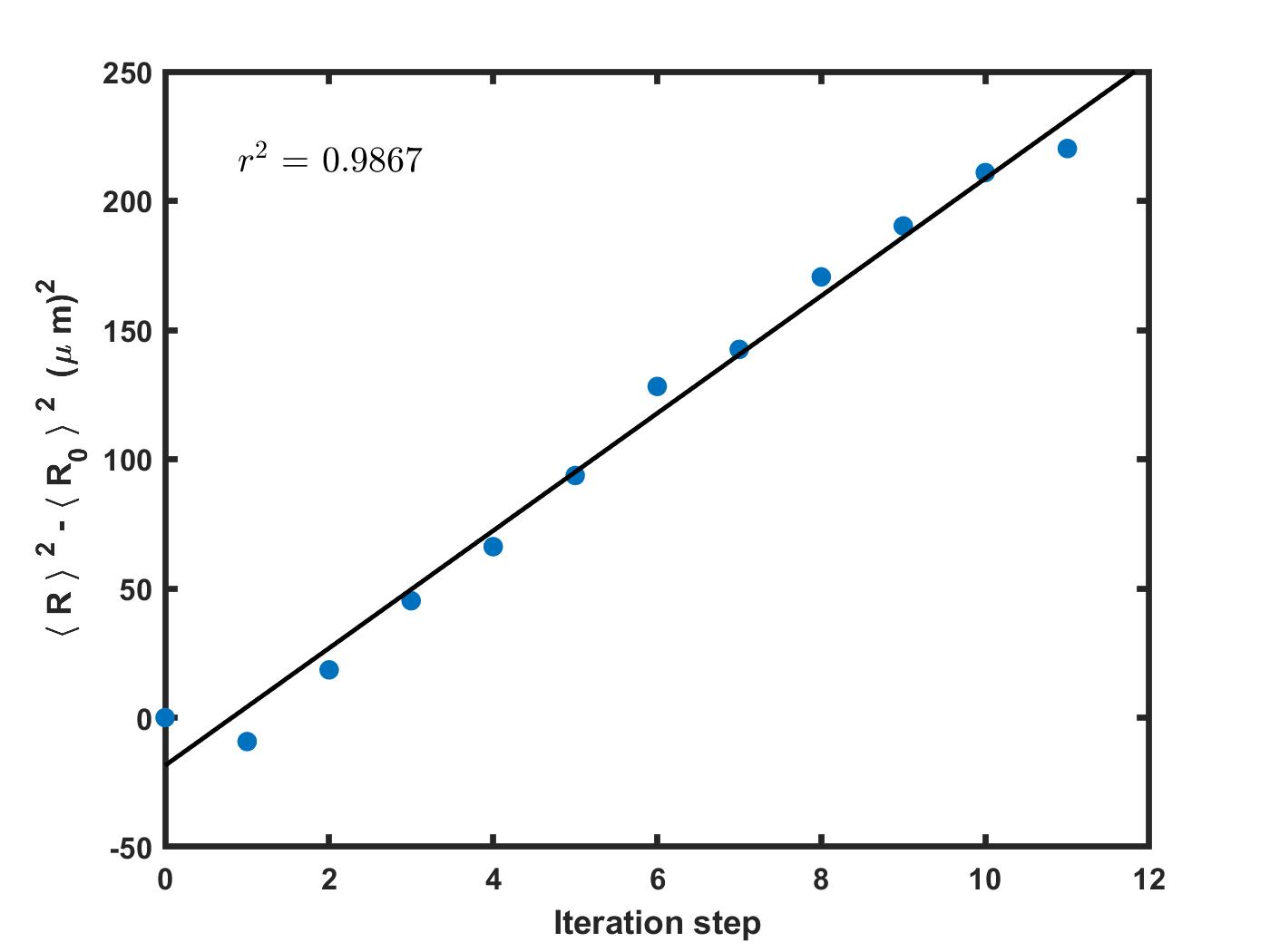}
\caption{\label{fig:R2} $\langle R \rangle ^2- \langle R_0 \rangle ^2$ as a function of iteration step. $r^2$ is the R-squared value from the linear fit of the data.}
\end{figure}

Burke and Turnbull \cite{burke1952recrystallization} found a linear relation between time and average grain size to the power n, where $2 \leq n \leq 4$, using an isotropic curvature-driven model for both 2D and 3D systems. The most commonly reported exponent for grain growth is n = 2.
In Figure \ref{fig:R2}, the difference between the 2nd power of average grain size $\langle R \rangle ^2$ and initial average grain size $\langle R_0 \rangle ^2$ of a simulation are plotted as a function of iteration step. 
The iteration steps can be interpreted as normalized time steps. 
The change in the square of the average grain size with time is well fit to a line, with a correlation coefficient ($R^2$) equal to 0.99.
Throughout this paper, grain size is calculated as $R = (\frac{3}{4\pi}V)^{1/3}$ with $V$ being the volume of the grain in voxels.  

Taken together, Figures \ref{fig:gsd} and \ref{fig:R2} indicate that the grain growth model produces results consistent with the classical theory. 
%%%%%%%%%%%%%%%%%%%%%
%%%%%%%%%%%%%%%%%%%%%
%%%%%%%%%%%%%%%%%%%%%
%%%%%%%%%%%%%%%%%%%%%
\subsection{Experimental data}
The 3D simulations described here are instantiated with observed Ni
microstructures and compared to a second observation of the same
sample at a later time.
Prior to the grain growth experiment at 800 $^{\circ}$C, the high-purity Ni was annealed at 950 $^{\circ}$C for 6 \cite{hefferan2009statistics}, which led to a completely recrystallized microstructure.  The X-ray data showed that the orientation spread within the grains was less than 0.1$^{\circ}$.  With this spread, the geometrically necessary dislocation density is less than 2.5$\times$10$^{11}$/m$^2$.  The estimated driving force for boundary migration from dislocations (the product of the dislocation energy per length and the dislocation density) is therefore on the order of 1 kJ/m$^3$.  The capillary driving force provided by 20$\times$10$^{-6}$ m grains is about 100 times larger.
The sample was measured at six instances in
time using near-field high energy X-ray diffraction microscopy \cite{hefferan2009statistics,hefferan2012measurement}
and was used to reconstruct the shapes and orientations of the grains
within six 3D volumes \cite{li2011imaging,hefferan2010tests}. Between each measurement, the
sample was annealed for $\approx 30$ min at $800^{\circ}$C. The details of the data
acquisition and interpretation have been described in previous
publications \cite{hefferan2009statistics,li2011imaging,hefferan2010tests}. For this work, we
segmented the grains and represented the microstructures as a set of
discrete voxels using DREAM.3D \cite{groeber2014dream} as described in \cite{bhattacharya2019}. The Ni microstructure has an abundance of
twin boundaries that have a significantly lower energy than other Ni
grain boundaries \cite{li2009relative,olmsted2009survey}. 
Obviously, such features cannot be reproduced by a simulation with isotropic boundaries properties. To create a microstructure with a narrower range of grain boundary energies that more closely approximates the isotropic grain growth model, all neighboring grains with the twin misorientation were merged to form a single grain using the merge twins function in DREAM.3D \cite{groeber2014dream}. 
After the twins were merged, the microstructures contained 600 to 900
grains made up of voxels with dimensions of $2.3 \times 2.3 \times 4.0 \, \mu$m. In
the initial state, there was an average of 7582 voxels per
grain.

%%%%%%%%%%%%%%%%%%%%%
%%%%%%%%%%%%%%%%%%%%%
%%%%%%%%%%%%%%%%%%%%%
%%%%%%%%%%%%%%%%%%%%%
\subsection{Data processing for comparison}

\subsubsection{Establishing the common volume.}
\label{Sec:ComVol}
The cylindrical Ni sample in the initial state is illustrated in Figure \ref{fig:ExSimWhole}(a), where the 920 grains are colored by orientation. 
The experimental volume was cylinder shaped with free surfaces around the periphery and orientations were sampled on a fixed grid. These geometric features determined the geometry of the simulation. 
The simulations volume is as rectangular parallalepiped, as illustrated by the frame in Figure \ref{fig:ExSimWhole}(a), where the transparent voxels do not belong to any grain.
The simulation directly uses the orientation map to create characteristic functions $\mathbf{1}_{\Sigma_i}$ as in Equation \ref{CharFn0}, and the phases $\Sigma_i^0$ are the input for iterations as in Algorithm \ref{alg:EO15}.
Thus, the computational grid of the simulation has the same size of the input microstructure, for example $369\times372\times62$ for anneal state 0.

Because of the cylindrical shape of the sample, a special procedure must be used to exclude the space that is inside of the computational grid but outside of the actual sample (transparent region in Figure \ref{fig:ExSimWhole}(a)). The easiest way to accomplish this is to manipulate the characteristic function of the transparent region $\Sigma_0$. By setting $\mathbf{1}_{\Sigma_0}=0, \text{if $x \in \Sigma_0$ }$ the contribution to $\psi_i$ ($i = 1, 2, ... 920$) from $\Sigma_0$ is always 0 in the convolution step in Algorithm \ref{alg:EO15}, and the boundary between $\Sigma_0$ and $\Sigma_i$ will not evolve during the iteration. Thus, the volume is preserved throughout the simulation.

To compare the microstructure in two different time steps, they must be in the same spatial reference frame and have the same volume.  Because non-identical fields of view were imaged at each time step, and the sample position was not perfectly reproduced, it is necessary to both translate the volume and crop it vertically to obtain a constant volume in a fixed frame of reference.  We first translated each volume until we identified the translation with the minimum disorientation between the two time steps. The larger volume was then cropped in the vertical direction so that both volumes had the same number of voxels at identical locations.

%%%%%%%%%%%%%%%%%%%%%
%%%%%%%%%%%%%%%%%%%%%
%%%%%%%%%%%%%%%%%%%%%
\subsubsection{Establishing the time scale.}
\label{Sec:TimeScl}
As the simulation proceeds, the average grain size increases with the square root of time, as illustrated in Figure \ref{fig:R2}. To compare with experiment, a criterion is needed to decide when to stop the simulation at an equivalent time. One was to use a voxel-by-voxel comparison of the grain identification (ID) number, counting all voxels that matched and taking the maximum of this as the point of best agreement. The second criterion was to determine the average distance that the boundaries moved in the simulation and match this with experiment.  The third criterion was to stop the simulation at the time when it had the closest average grain size as the experiments.  To decide which to use, we compared the number of grains in the simulated and experimental microstructure.  None of the methods were perfect, but the criterion based 
on the nearest average grain size led to grain counts in the simulation that were closest to the experiment.

%%%%%%%%%%%%%%%%%%%%%
%%%%%%%%%%%%%%%%%%%%%
\section{Results}
\label{sec:GG-results}
\begin{figure}[h!]
\centering
\includegraphics[width=1\textwidth]{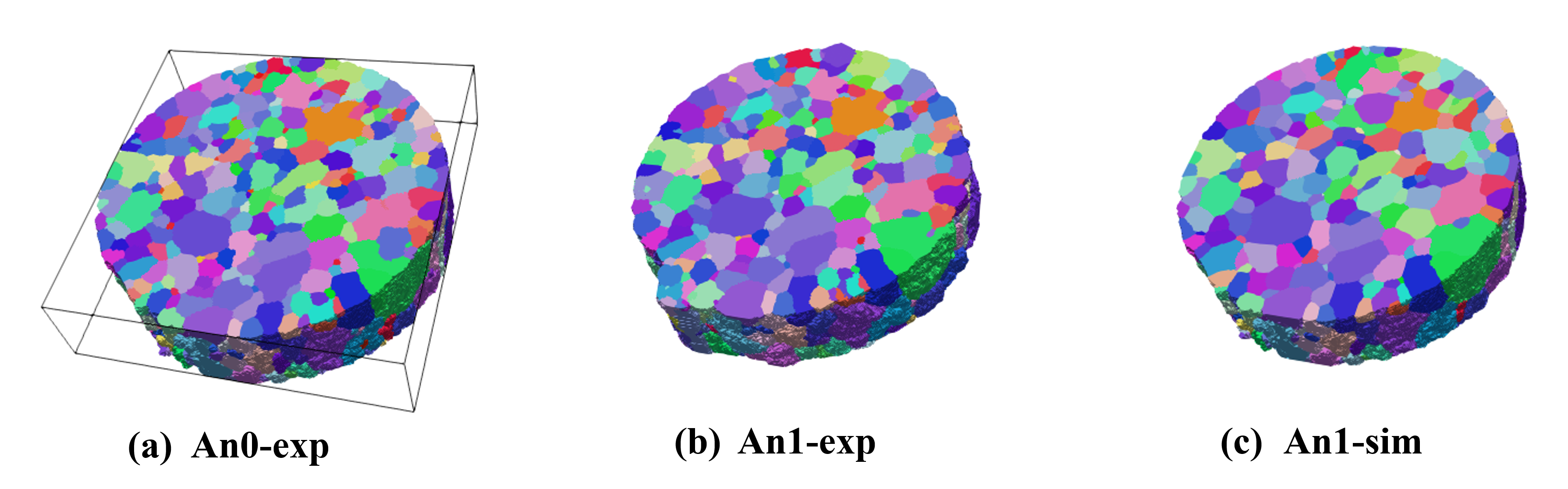}
\caption{\label{fig:ExSimWhole} (a) Experimentally measured microstructure of the initial anneal state of 920 grains and (b) experimental final state with 756 grains and (c) simulated microstructure of the final anneal state with 642 grains.}
\end{figure}

\begin{figure}[h!]
\centering
\includegraphics[width=0.9\textwidth]{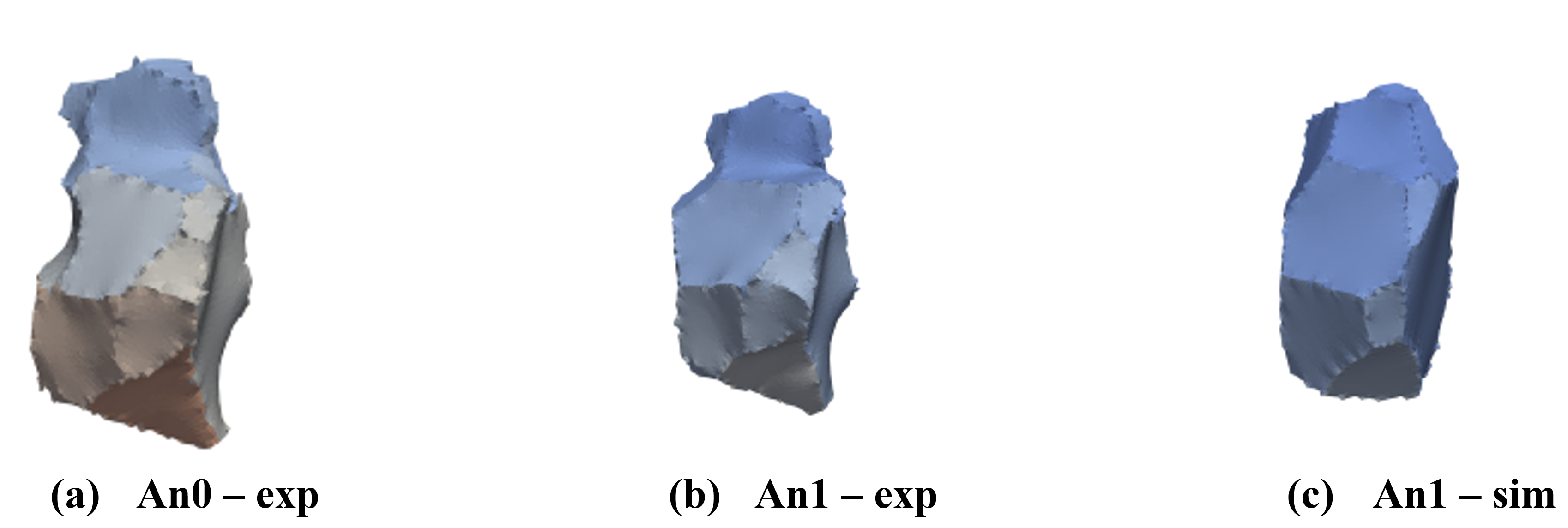}
\caption{\label{fig:ExSimBulk} A grain in experimentally measured microstructure of the (a) initial anneal state, (b) final anneal state and (c) the corresponding grain in the simulated microstructure of the final state. Grain faces are colored to make them distinguishable.}
\end{figure}

\begin{figure}[h!]
\centering
\includegraphics[width=1\textwidth]{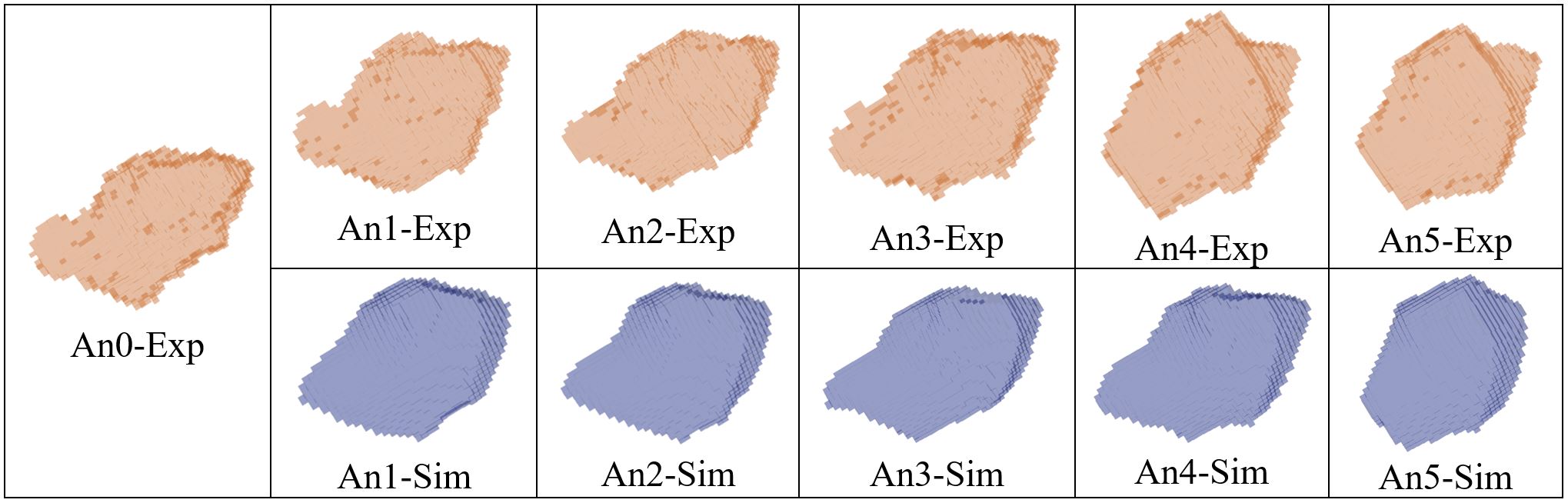}
\caption{\label{fig:grainAllStates}  Visualization of a bulk grain across all Anneal States in experiment and comparison with simulation of the same grain. The top row and bottom row shows experimental and simulation grains starting from An0 to An5 respectively.}
\end{figure}

Figure \ref{fig:ExSimWhole} 
(a) shows the experimentally measured microstructure of the initial anneal state of 920 grains 
and 
(b) the experimental final state microstructure with 756 grains
and 
(c) the simulated microstructure of the final anneal state with 642 grains. 
The average grain size in (a) is $26.0 \, \mu$m, in (b) is $26.8 \, \mu$m. and in (c) is $30.2 \, \mu$m. Between the grains that matched, the reconstructed initial experimental state has an average of 11 faces per grain, and the final state has 11 faces per grain while the simulation state has 9.6 faces per grain. Comparing the top surface between the two, the interfacial network of the simulated grains is smoother than the experiment. This will be addressed in detail later.
Figure \ref{fig:ExSimBulk} takes a closer look at the faces of a grain for the initial and final experimental state (a), (b) and the final simulation state (c). The number of grain faces changes in the simulation and experiment, but not in exactly the same way. While there are six anneal states, each state was used as an initial state to simulate the next one. A grain that does not contact any of the external surfaces has been visualized at all states as it evolves from Anneal State 0 to Anneal State 5 in Figure \ref{fig:grainAllStates}. This grain shrinks during each annealing interval in the experiment as well as simulation. 

\begin{figure}[h!]
\centering
\includegraphics[width=0.7\textwidth]{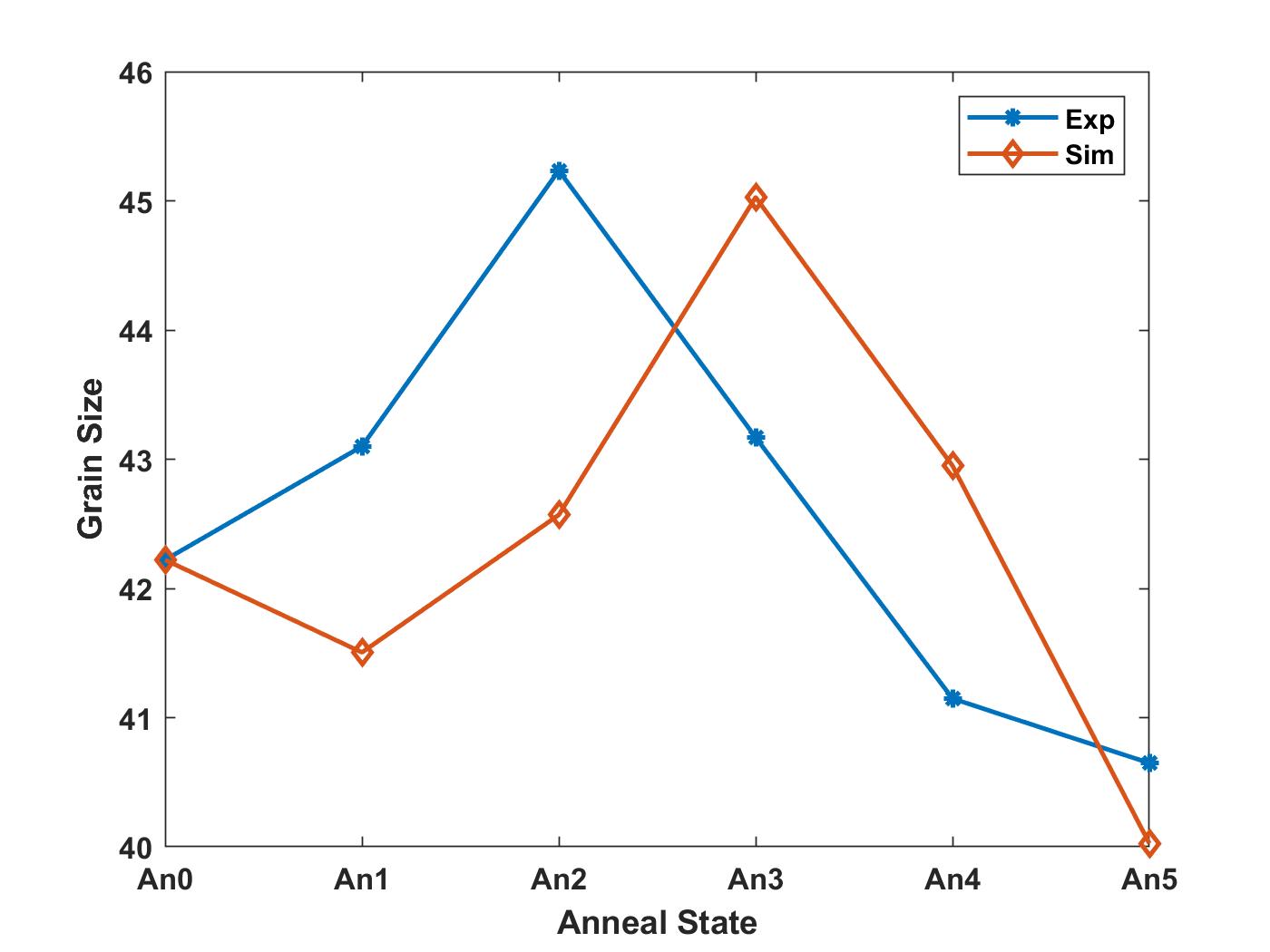}
\caption{\label{fig:ExSimGSOneG} The actual and simulated evolution of the grain size of the bulk grain in Figure \ref{fig:grainAllStates} at every anneal state.}
\end{figure}

The experimental and simulated changes in grain radius for the grain visualized in Figure \ref{fig:grainAllStates} have been plotted in Figure \ref{fig:ExSimGSOneG} as a function of the anneal states.  Starting at a radius of $42.2\, \mu$m for both, the experimental plot shows the grain grew until An2 and then shrank to $40.6\, \mu$m. While the simulated grain also shrank in the final state ($40 \, \mu$m), it did not follow the same size trajectory. For example, while both started from the same radius, the simulated grain shrank between An0-An1 but in the experiment it grew. Ultimately though, the grain shrank and the simulation could captured that.

\begin{figure}[h!]
\centering
\includegraphics[width=0.7\textwidth]{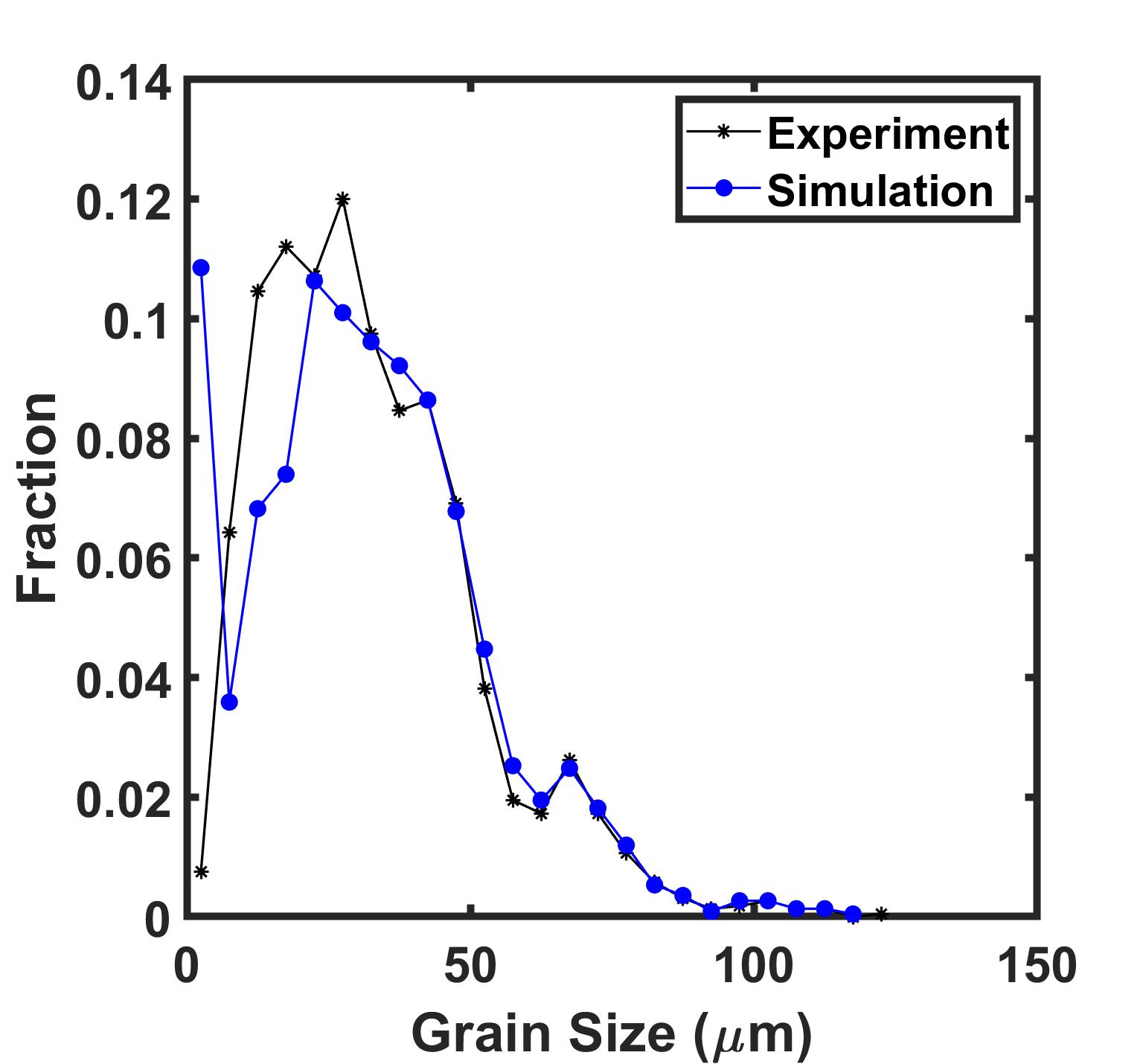}
\caption{\label{fig:ExSimGSD} The grain size distribution of observed and simulated microstructures for all the anneal states.}
\end{figure}

Figure \ref{fig:ExSimGSD} depicts the grain size distribution of observed and simulated microstructures for all the anneal states. While the simulation is mostly coincident with the actual grain size distribution, there are significant differences for the smaller grain sizes. The distribution shows that the simulated microstructure has a significantly higher fraction (0.11\%) of smaller grains (<$5 \, \mu$m) than in experiment (0.01\%). In other words, simulation is shrinking more grains than shrink in the experiment.

\begin{figure}[h!]
\centering
\includegraphics[width=.7\textwidth]{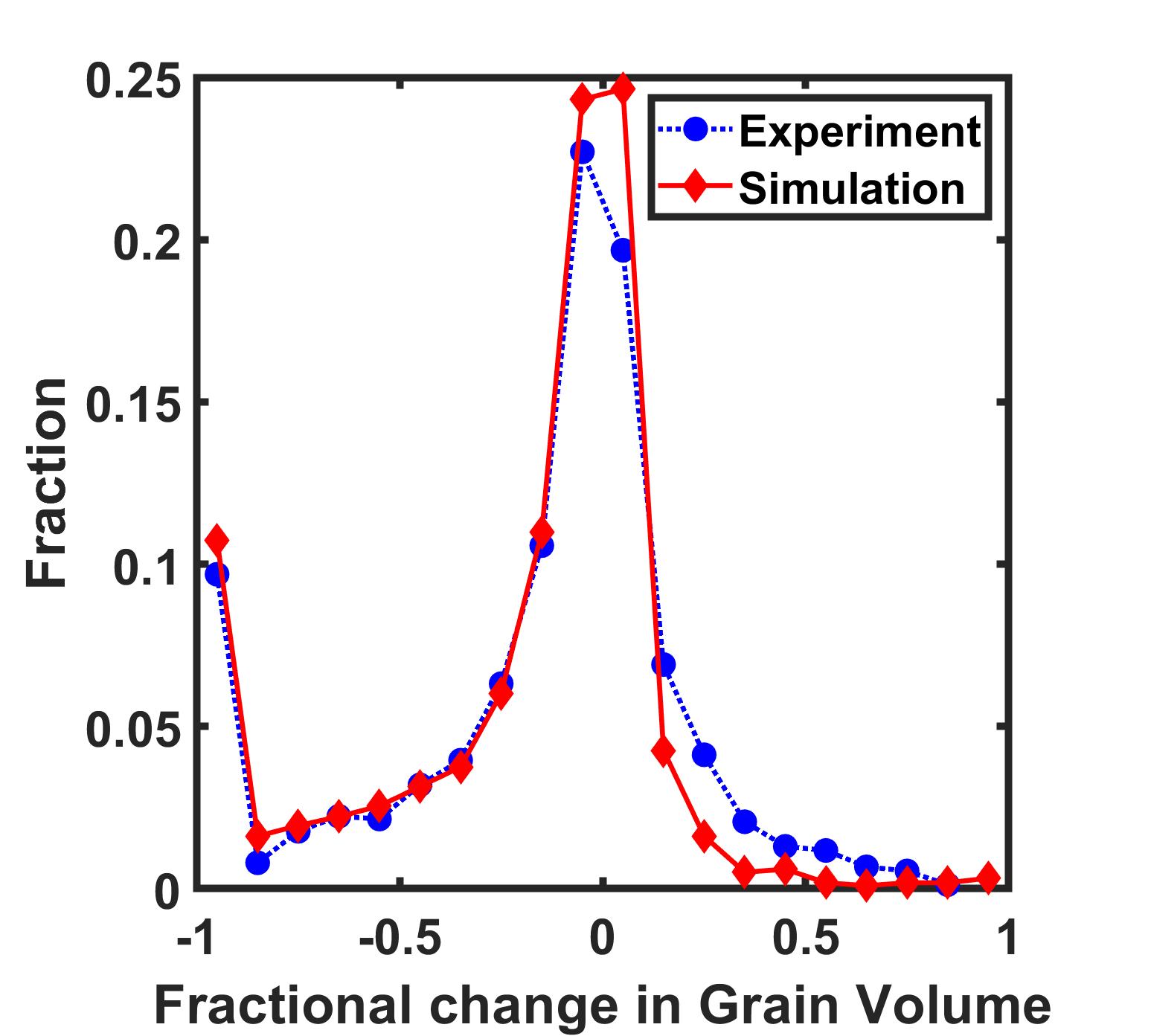}
\caption{\label{fig:ExSimFracVolD} Histogram of fractional change in volume for final state experimental and simulated grains. The markers show the bin centers and the lines connect bin centers to guide the eye.}
\end{figure}

Figure \ref{fig:ExSimFracVolD} plots the histogram of fractional change in volume for observed and simulated grains. 
The data includes all grains that matched from the initial anneal state to the final anneal state as well as the grains that were smaller than the average grain size that did not find a match and were assumed to have disappeared (fractional volume change of -1). 
It is seen that approximately the same fraction of grains have shrunk to zero in both experiment and simulation (~10\%).
In both simulation and experiment, the maximum of the distribution is for grains that have small volume changes near zero. Also, there are more grains in the experiment that increase in size as shown by the right tail of the distribution.

\begin{figure}[h!]
\centering
\includegraphics[width=.8\textwidth]{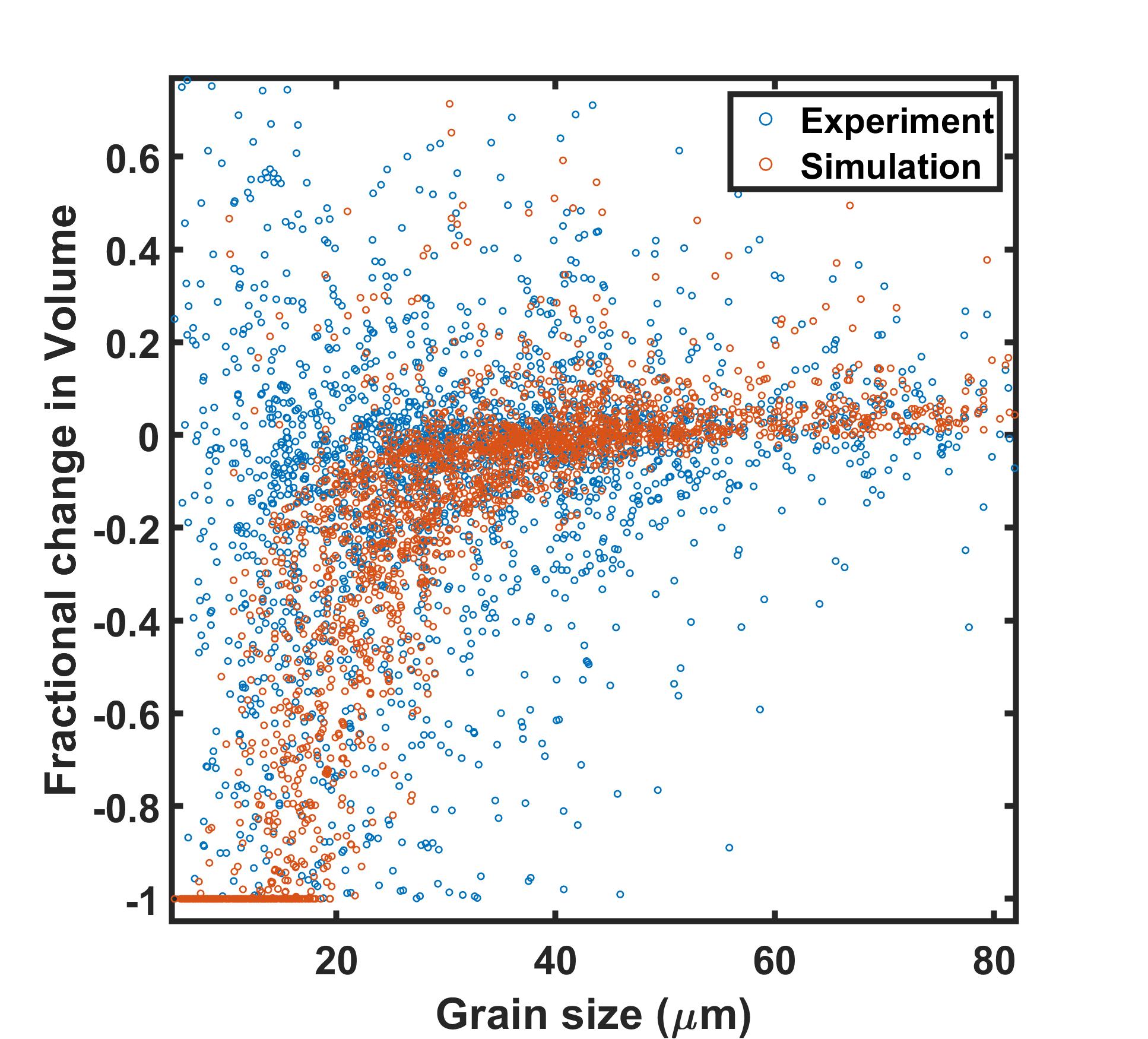}
\caption{\label{fig:ExSimFracVolGS} The fractional change in volume vs grain size in experimental and simulated grains.}
\end{figure}

To understand Figure \ref{fig:ExSimFracVolD} in more detail we consider Figure \ref{fig:ExSimFracVolGS}, which plots the fractional change in volume as a function of grain radius for both simulation and experiment. The simulation tends to shrink the small grains. Note that nearly all of the simulated grains in the size range of $5$ to $25 \,  \mu$m have negative volume changes while grains in the same size range in the experiment are equally probable to have positive or negative volume changes.
Counter-intuitively, most of the largest positive fractional volume changes occur for the small grains in the experiment.
When correlated with Figure \ref{fig:ExSimFracVolD}, it means that the lower extreme of fractional change in volume (<-0.6) corresponds to grain sizes less than $35 \, \mu$m. It also shows that in both experiment and simulation, as the grain size gets lager, the magnitude of fractional change is smaller. This suggested that the mismatch at the right tail of the distribution in Figure \ref{fig:ExSimFracVolD} is mainly caused by incorrect volume predictions for small grains. Notice that the grains that disappeared in the simulation are clustered in the small grain size region, whereas grains of the same sizes exhibit significant growth in the experiment. Based on Figure \ref{fig:ExSimFracVolGS}, we conclude that most of the differences between the volume changes in Figure \ref{fig:ExSimFracVolD} arise from errors in the volume predictions for the small grains. 

\begin{figure}[h!]
\centering
\includegraphics[width=0.8\textwidth]{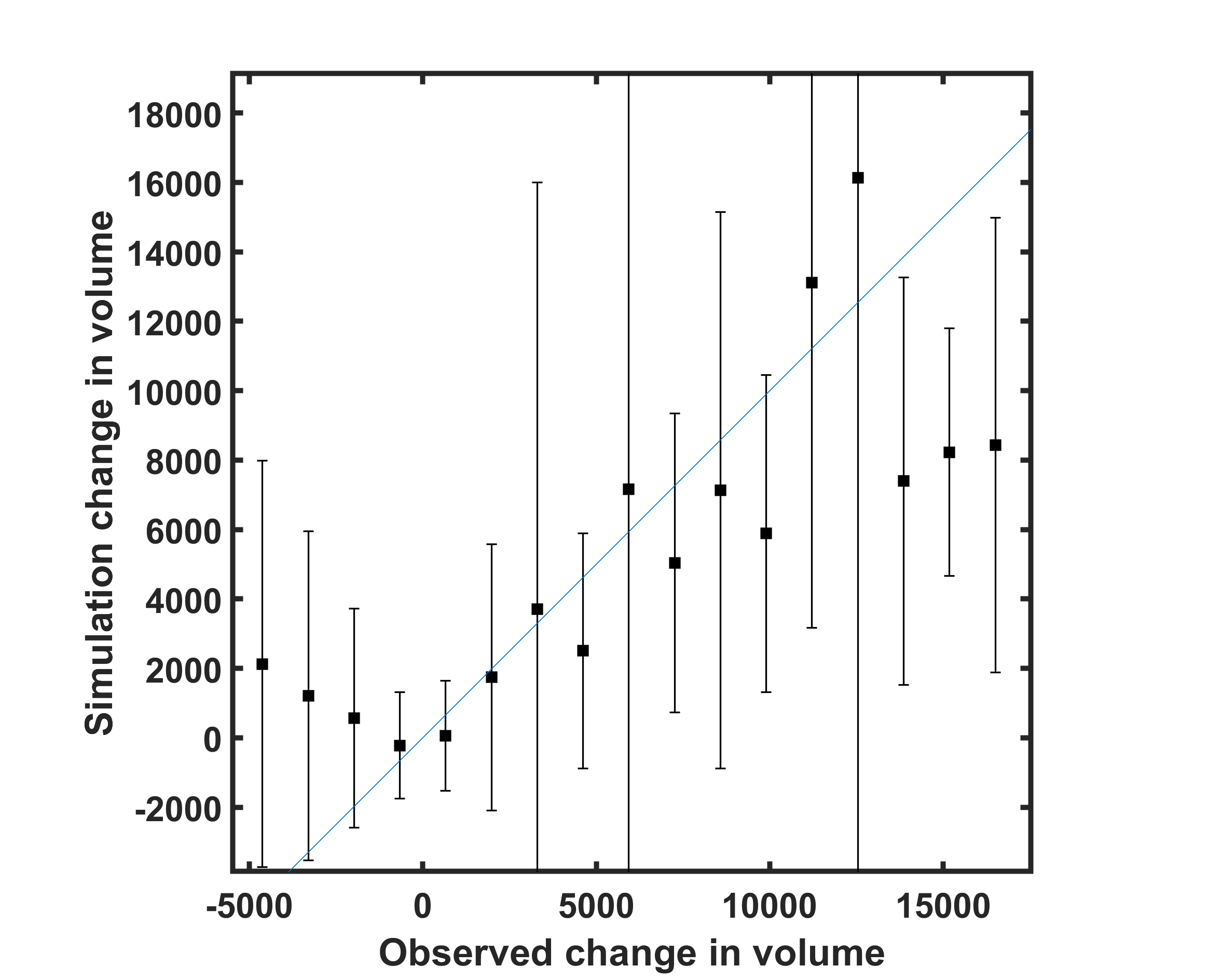}
\caption{\label{fig:VolPredict} Simulated volume change as a function of observed volume change in experiment for all matched grains.}
\end{figure}

Figure \ref{fig:VolPredict} compares simulated and observed volume changes in the same grains. 
The observed volume changes have been classified into discrete bins and the markers in the plot are the means of the values in each bin and the bars are the standard deviation. 
The negative extreme of the plot is much flatter; this is because grains that disappeared are not included. 
Therefore, the sum of the increased and decreased volumes is not zero. In other words, the volume is not conserved because of the over prediction of grain shrinkage. The plot shows a positive direct relationship between the experimentally observed and the simulated change in volume. When the experimental volume change is greater than zero, the simulated volume change also is positive, and we see a positive monotonic trend between the two. The slope of the trend is lower than the ideal $45^{\circ}$ line, meaning simulation under-predicts the grain growth. It was found that for all matched grains between each anneal state pair, considered together, the simulation has correctly predicted the sign of volume change 62\% of the time.

\begin{figure}[h!]
\centering
\includegraphics[width=0.65\textwidth]{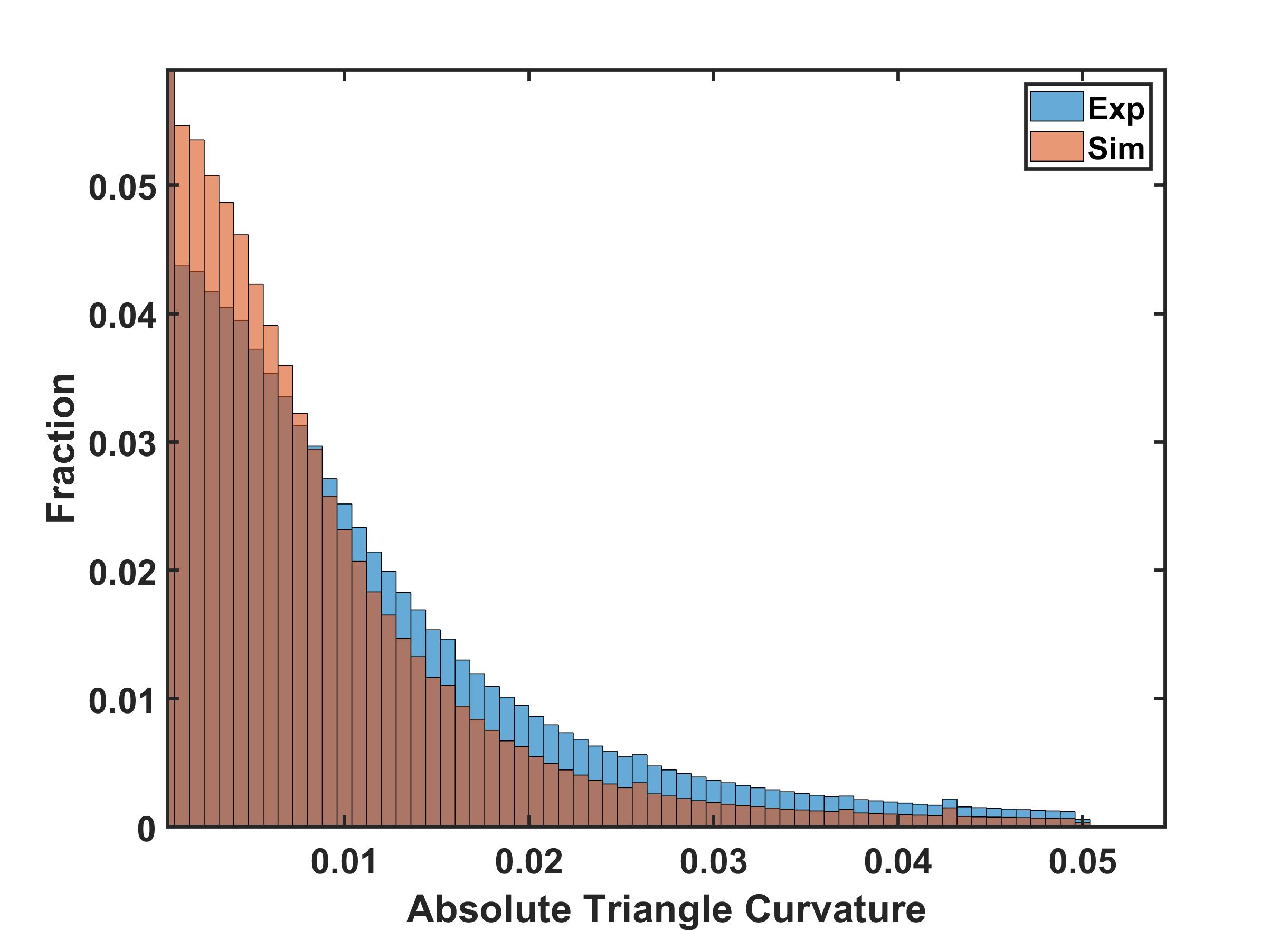}
\caption{\label{fig:ExSimAbsTriCur} Absolute triangle curvatures for all boundaries in the simulation and final experiment state.}
\end{figure}

Figure \ref{fig:ExSimAbsTriCur} shows a comparison between the unsigned triangle curvatures from the grain boundary mesh for all matched grain boundaries of the experiment and simulation. The y-axis displays the fraction of triangles in the simulation and experiment. The lower curvatures have been overpredicted by the simulation while it is underpredicting higher values of curvature. In other words, the simulation produces smaller grain boundary curvatures. Experimental curvatures have a higher fraction of triangles with curvatures greater than $0.008 \, \mu$m$^{-1}$ while simulation triangles have a higher fraction of lower curvature values (14\%) than experiment (12\%). This quantifies the visual impression one gets from comparing the experimental and simulated microstructures, that
the simulated microstructures have smoother interfaces.

\begin{figure}[h!]
\centering
\includegraphics[width=0.8\textwidth]{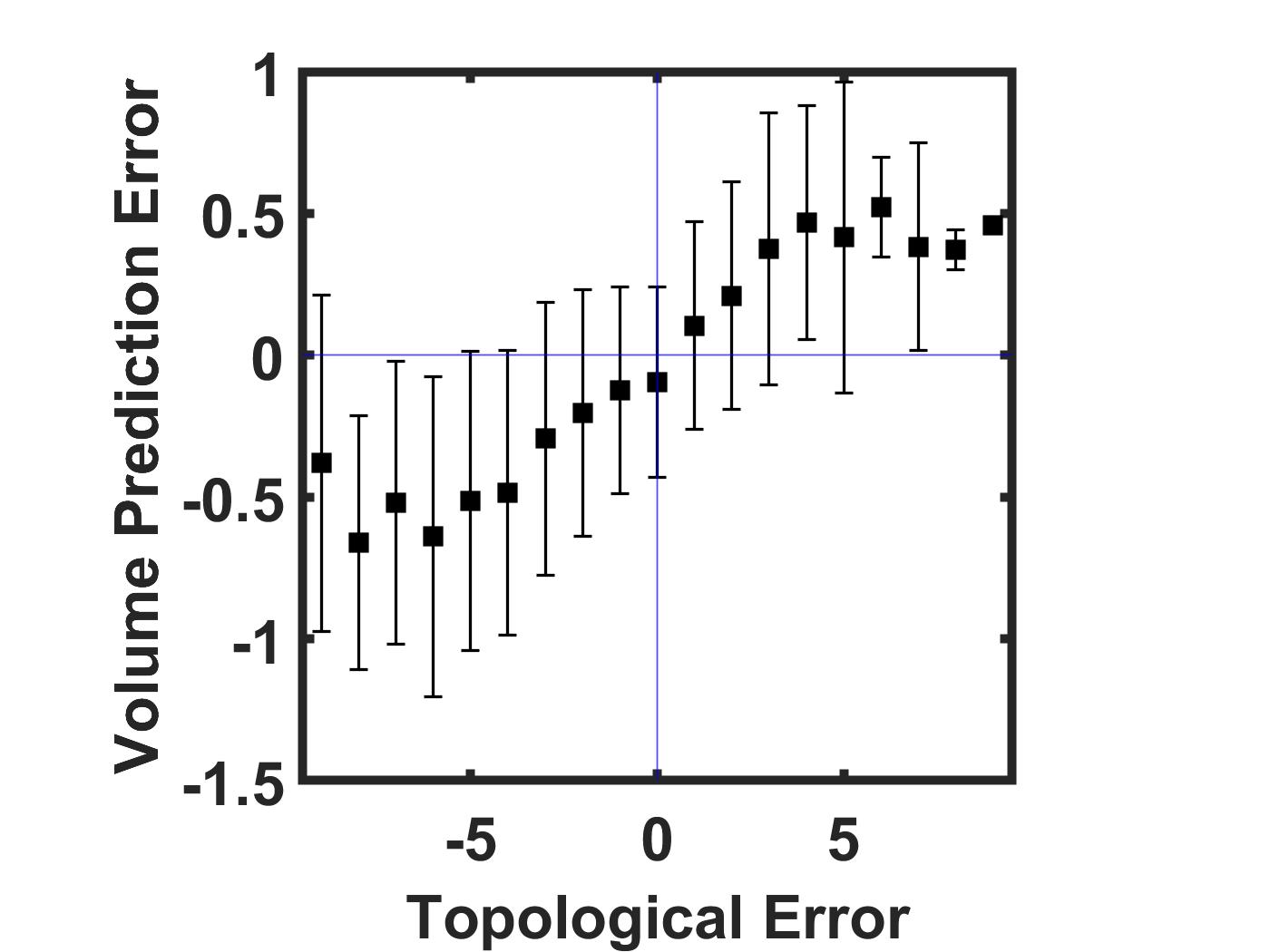}
\caption{\label{fig:ErrorTopoVol} The volume prediction error ($VPE$) as a function of topological error ($TE$). $VPE$ is the fractional difference in predicted and observed grain volume. $TE$ is the difference in grain face evolution between simulation and experiment.}
\end{figure}

One source of the volume error might be errors in the predicted topology of the grains and to test this hypothesis, we compare the volume prediction error ($VPE$) with the topological error ($TE$) for individual grains. $VPE$ and $TE$ are defined as follows:
\begin{align}
    \Delta N_s &= N_\text{sim} - N_\text{exp(initial)} \\
    \Delta N_e &= N_\text{exp(final)} - N_\text{exp(initial)} \\
    TE &= \Delta N_s - \Delta N_e \\
    VPE &= \frac{V_s - V_e}{V_e} \\
\end{align}
where $N_\text{exp(initial)}$ is the number of neighbors of a grain in the initial experiment state, $N_\text{exp(final)}$ is the number of neighbors of the same grain in the final experiment state and $N_\text{sim}$ is the number of neighbors of the same grain in the final simulation state. $VPE$ is the fractional difference in volume predicted by simulation of final anneal state ($V_s$) and experimental final state ($V_e$). $TE$ is the difference in $\Delta N$ for each grain between simulation and experiment. In other words, $TE$ is the error in predicting topological evolution by the simulation. Figure \ref{fig:ErrorTopoVol} plots the volume prediction error as a function of topological error. 
A low $VPE$ indicates a small difference between the final volume predicted and the actual final volume of the grain. A high $TE$ value means there is a large error in predicting the topological evolution of the grains. This plot is approximately linear and monotonically increasing. When $TE$ is close to zero, $VPE$ is also close to zero. As $TE$ increases, the error in the predicted volume also increases, in both positive and negative directions.

%%%%%%%%%%%%%%%%%%%%%
%%%%%%%%%%%%%%%%%%%%%
%%%%%%%%%%%%%%%%%%%%%
%%%%%%%%%%%%%%%%%%%%%
\section{Discussion}

\begin{figure}[h!]
\centering
\includegraphics[width=0.75\textwidth]{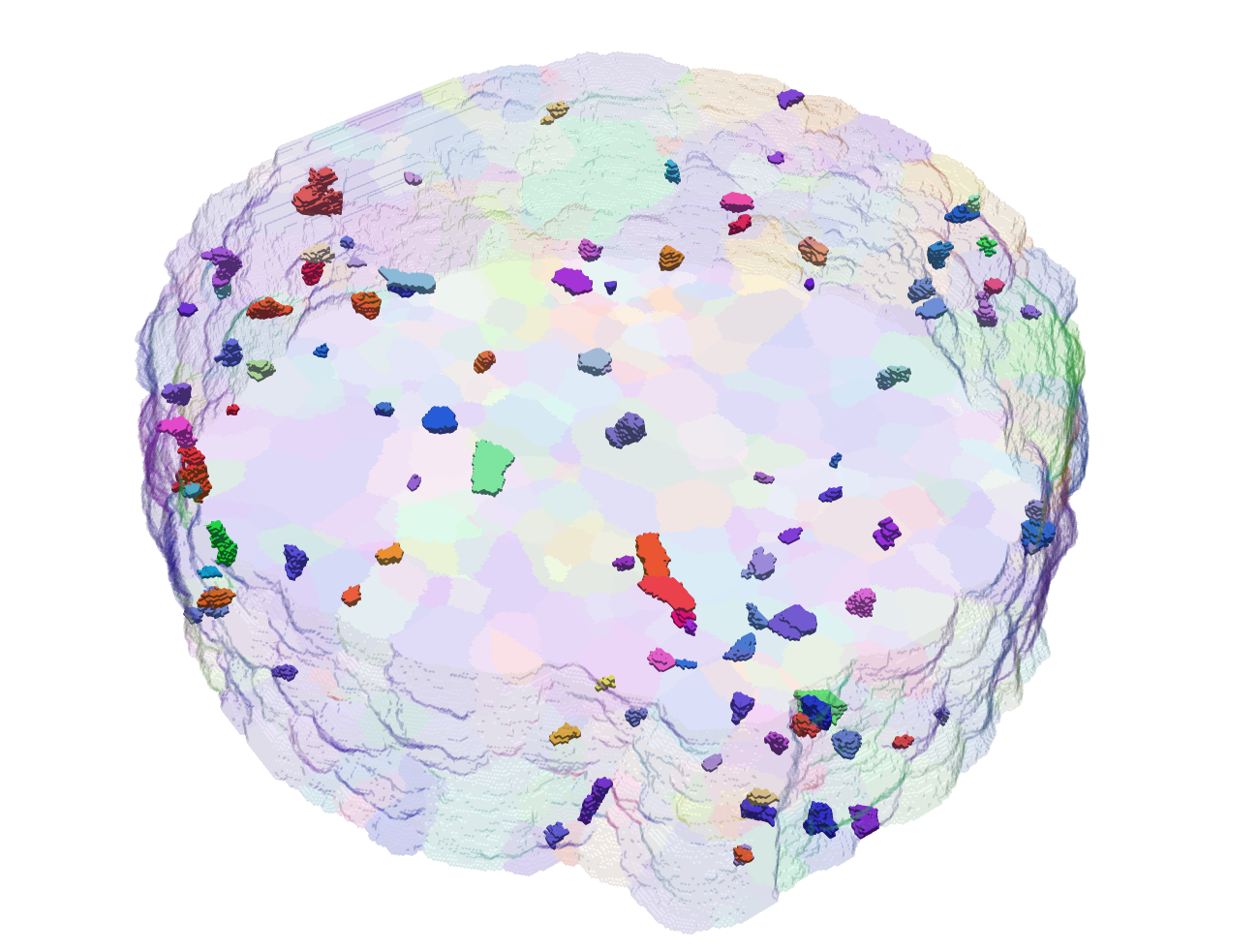}
\caption{\label{fig:DisapperG} Visualization of the 101 grains in An4 that shrunk to zero in An5 in simulation. These grains are colored by IPF coloring and are opaque while the rest of the microstructure is transparent.}
\end{figure}

Based on Figs. \ref{fig:ExSimFracVolD} and \ref{fig:ExSimFracVolGS}, one of the key features of the simulation that differ from the experiment is the behavior of the smallest grains.  
As an example, 
during the simulation of growth during one annealing period, 101 grains disappeared.  
These grains are illustrated in Fig. \ref{fig:DisapperG}, illustrating that they are relatively small and positioned randomly in the volume.  
Of these grains, 57 (47\%) also disappeared in the experiment.  Of the remaining grains, about 2/3 of them got smaller or did not change volume.  One supposes that eventually, these grains will also disappear.  However, the remaining grains (1/5 of the total) increased in volume, directly contrary to the simulations.  Whether a grain grows or not has more to do with its environment than its size \cite{bhattacharya2019,shen2019importance} and it has been shown that small grains can form and grow, provided they decrease the total interfacial energy \cite{lin2015observation}. 
However, even the grains undergoing a negative volume change in the experiment appear to shrink more slowly than in the simulation.  
This suggests the possibility of some physical process that retards the migration of boundaries around the smallest grains.  It does not seem likely that the decreased rate of shrinkage could be a result of the grain boundary energy anisotropy, which is not accounted for here.  
While a spectrum of grain boundary energies would alter the driving force and grain boundary migration rate, it is just as likely to increase the rate as decrease the rate, and the grain boundary energy is not expected to change with the grain size.  
One possibility is that as the grain boundary area decreases, the local concentration of segregated impurities \cite{gottstein2001grain} or boundary defects increases \cite{rabkin2020grain} and this influences migration kinetics.  However, this possibility cannot be tested with the available data.  
It was noted that the grain boundaries had, on average, lower curvatures than the experimental data set.  
This observation cannot explain the difference in the behavior of the shrinking grains, as reduced curvature should reduce the rate of grain volume change.  
Because the energies are anisotropic in the experiment, the curvature is not expected to be uniform as it is in the simulations.  The variations in local curvature in the experiment may be a result of grain boundary energy anisotropy or simply are an artifact of the experiment's finite spatial resolution.  In any case, it is difficult to predict how this would influence grain boundary migration; for the case of anisotropic grain boundary energy, the migration rate depends on the grain boundary stiffness and the anisotropy of the stiffness is greater than that of the grain boundary energy \cite{abdeljawad2018role}.  
However, grain boundary properties are not expected to change with size, so this does not explain the difference in the behavior of the small grains.

It was also found that errors in the volume change of each grain are tied to errors in topological changes.  In other words, if a grain's change in its number of neighbors is the same in the simulation and experiment, the volume change is more accurately predicted.  This is not too surprising.  Once the simulated and experimental grain have a different number of near neighbors, the sizes of the grain faces and grain face curvatures will be different, causing them to evolve differently.  The greater the difference in the number of neighbors, the larger the error in the volume change.  The source of these errors is difficult to identify, but one is certainly disappearance of more grains in the simulation.  Assuming the grains that disappear have a minimum of four neighbors, each grain that disappears in the simulation but not the experiment changes the number of neighbors of at least four grains.  For the example discussed above, where 64 more grains disappear in the simulation than in the experiment, this changes the number of neighbors of as many as 256 grains. 

There have been a number of recent grain growth simulations that assume anisotropic grain boundary energies \cite{miyoshi2021large,salama2020role,gruber2005effect,gruber2009misorientation}.
It is envisioned that if realistic models for the grain boundary energy anisotropy are used, they will better predict the evolution of the microstructure.  
However, the current results suggest that it might be necessary to include a size dependent migration model because grains that disappear too quickly change the environments of other grains and this leads to errors in the prediction of volume changes. 
In future work, the isotropic TD scheme used here will be extended to include anisotropic surface energy and mobility data from experiment \cite{esedoglu2017kernels}. This platform will make it possible to explore different forms of the anisotropic kernel that will best model the anisotropic grain growth in Ni.

\section{Conclusions}

Experimentally observed Ni microstructures, at six time steps, were compared in a grain-by-grain fashion to the results of isotropic grain growth simulations. This comparison led to the following conclusions.
The simulation predicts the correct sign of the volume change for only 62\% of the grains. The errors are the most pronounced for the smaller grains, for which shrinkage is over predicted and growth is under predicted.
Improved grain growth models might need to incorporate migration kinetics influenced by grain size. 
The grain boundary curvatures in the simulation are systematically lower than in the experimental observations. 
Volume prediction errors are correlated to errors in predicting topological changes. When the simulation captures the topological changes correctly, it can predict the grain volume change accurately as well.
%%%%%%%%%%%%%%%%%%%%%
%%%%%%%%%%%%%%%%%%%%%
%%%%%%%%%%%%%%%%%%%%%
%%%%%%%%%%%%%%%%%%%%%	

\section*{Software Availability}

A version of the code developed for this work is available at \\
\url{https://github.com/JadeXiaoyaoPeng/GrainGrowth_TD_iso}

\begin{acknowledgments}
    This work was supported by the National Science Foundation under DMREF grants 1628994 and 2118945.
    We acknowledge NSF for XSEDE computing resources provided by Pittsburgh Supercomputing Center.
    This research used resources of the Advanced Photon Source, a U.S. Department of Energy (DOE) Office of Science User Facility operated for the DOE Office of Science by Argonne National Laboratory under Contract No. DE-AC02-06CH11357.
\end{acknowledgments}

%%%%%%%%%%%%%%%%%%%%%
%%%%%%%%%%%%%%%%%%%%%
%%%%%%%%%%%%%%%%%%%%%
%%%%%%%%%%%%%%%%%%%%%	


\end{document}

%% file: preamble.tex
\usepackage{isomath}
\usepackage{amsmath,amsthm}
\usepackage{amsbsy}
\usepackage{amssymb}
\usepackage{amscd}
\usepackage{amsfonts}
\usepackage{stmaryrd}
\usepackage{siunitx}
\usepackage{euscript}
\usepackage[utf8]{inputenc}
\usepackage[T1]{fontenc}
\usepackage{newtxtext} 
\everymath{\displaystyle}
\usepackage{exscale}

\usepackage{graphicx}
\usepackage{boxedminipage}
\usepackage{calc}
\usepackage[usenames,dvipsnames]{xcolor}
\graphicspath{ {imagedir/} }
\usepackage[caption=false,justification=centerlast]{subfig}
\usepackage{setspace}
\usepackage{enumitem}
\setitemize{noitemsep,topsep=0pt,parsep=0pt,partopsep=0pt}
\setenumerate{noitemsep,topsep=0pt,parsep=0pt,partopsep=0pt}
\setdescription{noitemsep,topsep=0pt,parsep=0pt,partopsep=0pt}

\usepackage[colorinlistoftodos, color=green!40,prependcaption]{todonotes}

\usepackage{hyperref}
\usepackage[normalem]{ulem}

\usepackage[small]{titlesec}

\titlespacing*{\section}{0pt}{12pt plus 4pt minus 2pt}{2pt plus 2pt minus 2pt}
\titlespacing*{\subsection}{0pt}{12pt plus 4pt minus 2pt}{2pt plus 2pt minus 2pt}
\titlespacing*\subsubsection{0pt}{12pt plus 4pt minus 2pt}{2pt plus 2pt minus 2pt}
\titlespacing*\paragraph{0pt}{12pt plus 4pt minus 2pt}{2pt plus 2pt minus 2pt}

\makeatletter

    \renewcommand*{\p@subsection}{}
    
    \renewcommand*{\p@subsubsection}{}
\makeatother

\setuptodonotes{inline}
\input{definitions}

%% file: definitions.tex
\usepackage{isomath}
\usepackage{amsmath}
\usepackage{amssymb}
\usepackage{amscd}
\usepackage{amsfonts}

\theoremstyle{definition}

\newcommand{\bfn}{{\mathbold n}}

\newcommand{\bfv}{{\mathbold v}}

\newcommand{\bfx}{{\mathbold x}}